\documentclass[
  twocolumn,
  prd,
  amssymb,
  preprintnumbers,superscriptaddress,
nofootinbib]{revtex4-1}

\pdfoutput=1
\usepackage{paralist}
\usepackage{graphicx}
\usepackage{enumitem}
\usepackage{latexsym}
\usepackage{amsfonts}
\usepackage{amssymb}
\usepackage{xcolor}
\usepackage[export]{adjustbox}
\usepackage{amsmath}
\usepackage[thinlines]{easytable}
\usepackage{array}
\usepackage{slashed}
\usepackage{dcolumn}
\usepackage{verbatim}
\usepackage{float}
\usepackage{multirow}
\usepackage{xspace}
\usepackage[normalem]{ulem}
\usepackage[
pdfauthor={Djuna Croon}]{hyperref}
\usepackage{tabularx}
\usepackage{lettrine}
\usepackage{setspace}
\usepackage{cleveref}
\usepackage{adjustbox}

\input Zallman.fd

\LettrineTextFont{\itshape}

\newcommand{\beq}{\begin{equation}}
\newcommand{\eeq}{\end{equation}}
\newcommand{\bea}{\begin{eqnarray}}
\newcommand{\eea}{\end{eqnarray}}

\newcommand{\refjnl}[1]{{\rm#1}}

\def\apj{\refjnl{ApJ}}                 
\def\aap{\refjnl{A\&A}}                

\usepackage{pdfbase}[2017/03/16]
\usepackage{xparse,ocgbase}
\usepackage{xcolor,calc}
\usepackage{tikzpagenodes,linegoal}
\usetikzlibrary{calc}
\usepackage{tcolorbox}

\ExplSyntaxOn
\let\tpPdfLink\pbs_pdflink:nn
\let\tpPdfAnnot\pbs_pdfannot:nnnn\let\tpPdfLastAnn\pbs_pdflastann:
\let\tpAppendToFields\pbs_appendtofields:n
\def\tpPdfXform{\pbs_pdfxform:nnnnn{1}{1}{}{}}
\let\tpPdfLastXform\pbs_pdflastxform:
\let\cListSet\clist_set:Nn\let\cListItem\clist_item:Nn
\ExplSyntaxOff

\usepackage{pdfbase}[2017/03/16]
\usepackage{xparse,ocgbase}
\usepackage{xcolor,calc}
\usepackage{tikzpagenodes,linegoal}
\usetikzlibrary{calc}
\usepackage{tcolorbox}

\ExplSyntaxOn
\let\tpPdfLink\pbs_pdflink:nn
\let\tpPdfAnnot\pbs_pdfannot:nnnn\let\tpPdfLastAnn\pbs_pdflastann:
\let\tpAppendToFields\pbs_appendtofields:n
\def\tpPdfXform{\pbs_pdfxform:nnnnn{1}{1}{}{}}
\let\tpPdfLastXform\pbs_pdflastxform:
\let\cListSet\clist_set:Nn\let\cListItem\clist_item:Nn
\ExplSyntaxOff

\makeatletter
\NewDocumentCommand{\tooltip}{%
  ssssO{\ifdefined\@linkcolor\@linkcolor\else blue\fi}mO{yellow!20}mO{0pt,0pt}%
}{{%
  \leavevmode%
  \IfBooleanT{#2}{%
    \ocgbase@new@ocg{tipOCG.\thetcnt}{%
      /Print<</PrintState/OFF>>/Export<</ExportState/OFF>>%
    }{false}%
    \xdef\tpTipOcg{\ocgbase@last@ocg}%
    \ocgbase@add@ocg@to@radiobtn@grp{tool@tips}{\ocgbase@last@ocg}%
  }%
  \tpPdfLink{%
    \IfBooleanTF{#4}{%
      /Subtype/Link/Border[0 0 0]/A <</S/SetOCGState/State [/Toggle \tpTipOcg]>>
    }{%
      /Subtype/Screen%
      /AA<<%
        \IfBooleanTF{#3}{%
          /E<</S/SetOCGState/State [/Toggle \tpTipOcg]>>%
        }{%
          \IfBooleanTF{#2}{%
            /E<</S/SetOCGState/State [/ON \tpTipOcg]>>%
            /X<</S/SetOCGState/State [/OFF \tpTipOcg]>>%
          }{
            \IfBooleanTF{#1}{%
              /E<</S/JavaScript/JS(%
                var fd=this.getField('tip.\thetcnt');%
                if(typeof(click\thetcnt)=='undefined'){%
                  var click\thetcnt=false;%
                  var fdor\thetcnt=fd.rect;var dragging\thetcnt=false;%
                }%
                if(fd.display==display.hidden){%
                  fd.delay=true;fd.display=display.visible;fd.delay=false;%
                }else{%
                  if(!click\thetcnt&&!dragging\thetcnt){fd.display=display.hidden;}%
                  if(!dragging\thetcnt){click\thetcnt=false;}%
                }%
                this.dirty=false;%
              )>>%
            }{%
              /E<</S/JavaScript/JS(%
                var fd=this.getField('tip.\thetcnt');%
                if(typeof(click\thetcnt)=='undefined'){%
                  var click\thetcnt=false;%
                  var fdor\thetcnt=fd.rect;var dragging\thetcnt=false;%
                }%
                if(fd.display==display.hidden){%
                  fd.delay=true;fd.display=display.visible;fd.delay=false;%
                }%
               this.dirty=false;%
              )>>%
              /X<</S/JavaScript/JS(%
                if(!click\thetcnt&&!dragging\thetcnt){fd.display=display.hidden;}%
                if(!dragging\thetcnt){click\thetcnt=false;}%
                this.dirty=false;%
              )>>%
            }%
            /U<</S/JavaScript/JS(click\thetcnt=true;this.dirty=false;)>>%
            /PC<</S/JavaScript/JS (%
              var fd=this.getField('tip.\thetcnt');%
              try{fd.rect=fdor\thetcnt;}catch(e){}%
              fd.display=display.hidden;this.dirty=false;%
            )>>%
            /PO<</S/JavaScript/JS(this.dirty=false;)>>%
          }%
        }%
      >>%
    }%
  }{{\color{#5}#6}}%
  \sbox\tiptext{%
    \IfBooleanT{#2}{%
      \ocgbase@oc@bdc{\tpTipOcg}\ocgbase@open@stack@push{\tpTipOcg}}%
    \tcbox[colframe=black,colback=#7,size=fbox,arc=1ex,sharp corners=southwest]{#8}%
    \IfBooleanT{#2}{\ocgbase@oc@emc\ocgbase@open@stack@pop\tpNull}%
  }%
  \cListSet\tpOffsets{#9}%
  \edef\twd{\the\wd\tiptext}%
  \edef\tht{\the\ht\tiptext}%
  \edef\tdp{\the\dp\tiptext}%
  \tipshift=0pt%
  \IfBooleanTF{#2}{%
    \setlength\whatsleft{\linegoal}%
  }{%
    \measureremainder{\whatsleft}%
  }%
  \ifdim\whatsleft<\dimexpr\twd+\cListItem\tpOffsets{1}\relax%
    \setlength\tipshift{\whatsleft-\twd-\cListItem\tpOffsets{1}}\fi%
  \IfBooleanF{#2}{\tpPdfXform{\tiptext}}%
  \raisebox{\heightof{#6}+\tdp+\cListItem\tpOffsets{2}}[0pt][0pt]{%
    \makebox[0pt][l]{\hspace{\dimexpr\tipshift+\cListItem\tpOffsets{1}\relax}%
    \IfBooleanTF{#2}{\usebox{\tiptext}}{%
      \tpPdfAnnot{\twd}{\tht}{\tdp}{%
        /Subtype/Widget/FT/Btn/T (tip.\thetcnt)%
        /AP<</N \tpPdfLastXform>>%
        /MK<</TP 1/I \tpPdfLastXform/IF<</S/A/FB true/A [0.0 0.0]>>>>%
        /Ff 65536/F 3%
        /AA <<%
          /U <<%
            /S/JavaScript/JS(%
              var fd=event.target;%
              var mX=this.mouseX;var mY=this.mouseY;%
              var drag=function(){%
                var nX=this.mouseX;var nY=this.mouseY;%
                var dX=nX-mX;var dY=nY-mY;%
                var fdr=fd.rect;%
                fdr[0]+=dX;fdr[1]+=dY;fdr[2]+=dX;fdr[3]+=dY;%
                fd.rect=fdr;mX=nX;mY=nY;%
              };%
              if(!dragging\thetcnt){%
                dragging\thetcnt=true;Int=app.setInterval("drag()",1);%
              }%
              else{app.clearInterval(Int);dragging\thetcnt=false;}%
              this.dirty=false;%
            )%
          >>%
        >>%
      }%
      \tpAppendToFields{\tpPdfLastAnn}%
    }%
  }}%
  \stepcounter{tcnt}%
}}
\makeatother
\newsavebox\tiptext\newcounter{tcnt}
\newlength{\whatsleft}\newlength{\tipshift}
\newcommand{\measureremainder}[1]{%
  \begin{tikzpicture}[overlay,remember picture]
    \path let \p0 = (0,0), \p1 = (current page.east) in
      [/utils/exec={\pgfmathsetlength#1{\x1-\x0}\global#1=#1}];
  \end{tikzpicture}%
}

\allowdisplaybreaks

\begin{document}

\title{
Anomaly Detection to identify Transients in LSST Time Series Data
}

\author{Miguel Crispim Romao}
\email{miguel.romao@durham.ac.uk}
\affiliation{Institute for Particle Physics Phenomenology, Department of Physics, Durham University, Durham DH1 3LE, U.K.}
\author{Djuna Croon} \email{djuna.l.croon@durham.ac.uk}
\affiliation{Institute for Particle Physics Phenomenology, Department of Physics, Durham University, Durham DH1 3LE, U.K.}
\author{Daniel Godines} \email{godines@nmsu.edu}
\affiliation{New Mexico State University, 1780 E University Ave, Las Cruces, NM 88003, USA.}

\date{\today}

\begin{abstract}
We introduce a novel approach to detecting microlensing events and other transients in light curves, utilising the isolation forest (iForest) algorithm for anomaly detection. Focusing on the Legacy Survey of Space and Time (LSST) by the Vera C. Rubin Observatory, we show that an iForest trained on \emph{signal-less} light curves can efficiently identify microlensing events by different types of dark objects and binaries, as well as variable stars. We further show that the iForest has real-time applicability through a drip-feed analysis, demonstrating its potential as a valuable tool for LSST alert brokers to efficiently prioritise and classify transient candidates for follow-up observations.
\end{abstract}

\preprint{IPPP/25/15}

\maketitle

\section{Introduction}
The upcoming Legacy Survey of Space and Time (LSST) by the Vera C. Rubin Observatory will redefine time-domain astrophysics.
LSST’s wide-field, deep-imaging strategy will capture an unprecedented volume of light curve data, revealing a diverse array of transient phenomena. Among these are gravitational microlensing events, including by potential dark matter  (DM) candidates \cite{LSSTDarkMatterGroup:2019mwo}, which are characterised by the transient brightening of a background star due to the gravitational field of an intervening object.  
A central challenge in leveraging LSST’s dataset is the early recognition of such transient signals, which is critical for initiating timely follow-up observations. 

Recent advancements in outlier and anomaly detection for transient surveys have leveraged machine learning techniques to identify rare or novel astrophysical events in real-time (e.g. \cite{Muthukrishna:2021msd,Gupta:2024haf,Aleo:2024bct}). In preparation for LSST, efforts have been made to develop classification algorithms tailored to its alert stream \cite{Soraisam_2020}, as well as deep-learning models capable of identifying transients in real time \cite{Shah:2025eik}. 

In previous work \cite{godines2019machine}, one of us introduced the use of a machine learning classifier to search for microlensing in wide-field surveys with low cadence data. In a recent work \cite{CrispimRomao:2024nbr}, two of us adapted this technique to search for different dark objects, focusing on two representative classes: boson stars, with flat density profiles leading to characteristic light curves with caustic peaks, and Navarro-Frenk-White (NFW) subhalo profiles, which are more sharply peaked and do not exhibit such features in their light curves.

\begin{figure}[ht]
  \centering
  \includegraphics[width=.4\textwidth]{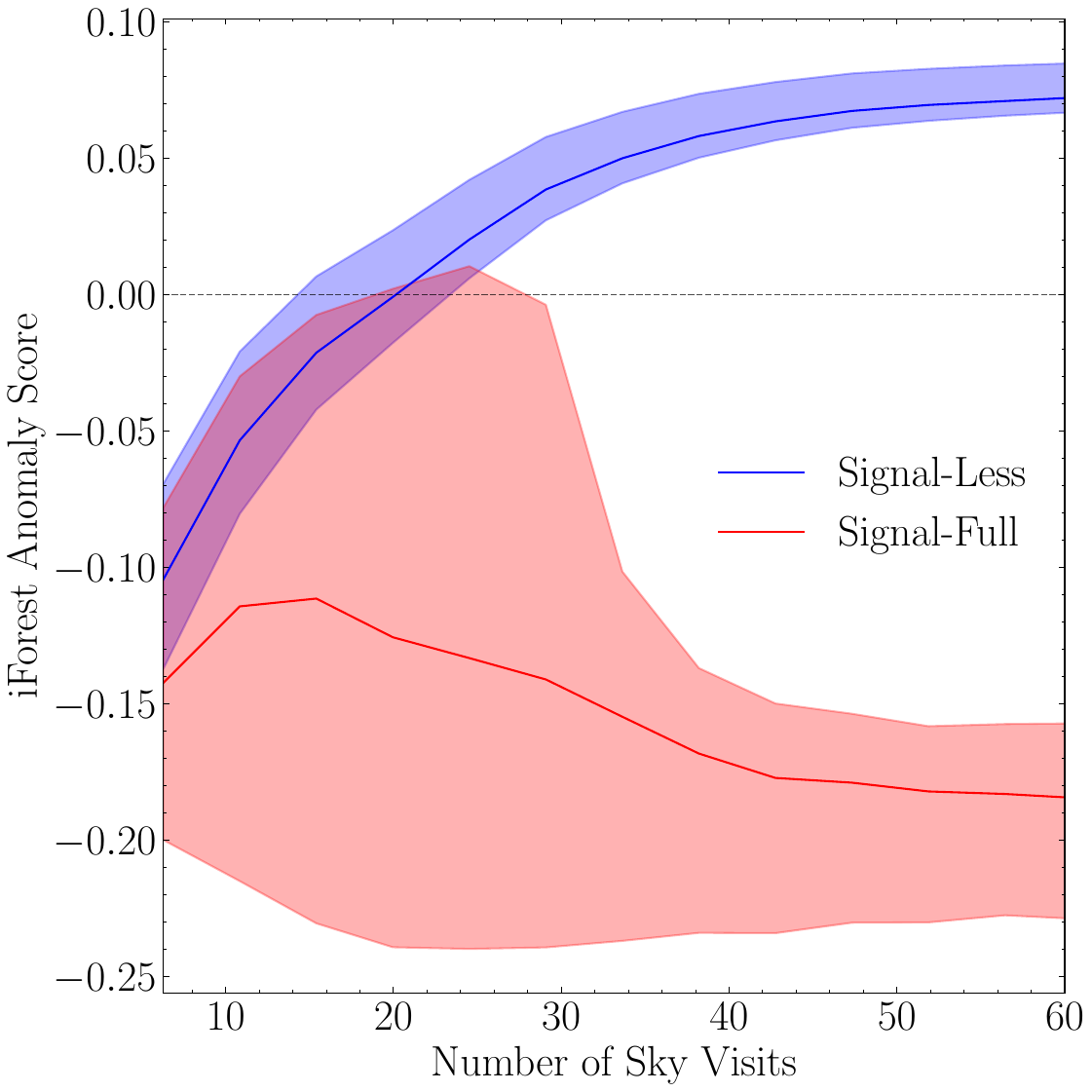}
  \caption{
  Comparison of anomaly scores for light curves with (this includes microlensing events with several lens profiles (point-like, boson star, NFW subhaloes, and binaries) and without signals. 
  The solid lines show mean bin values, while shaded regions represent the 25th–75th percentile range. 
  }
  \label{fig:main}
\end{figure}

In this work, we pioneer the use of
anomaly detection by means of an isolation forest (iForest) to identify microlensing events and other transients in LSST light curves. 
We utilise the cadence simulations provided by the Rubin Observatory to accurately reproduce the transient signals that will be observed during the LSST. These Operational Simulations (``OpSims", \cite{Bianco+22})\footnote{https://www.lsst.org/content/charge-survey-cadence-optimisation-committee-scoc} provide simulated observations using the Rubin scheduler, thus allowing the community to assess different survey strategies so as to optimise survey parameters for maximum scientific output, including the cadence and sky coverage of the survey.
The defining innovation in this work is that we also use ``OpSims'' to simulate a \emph{signal-less} or ``constant'' class, which we use to train our iForest model on.
We find that an iForest trained on signal-less light curves can effectively identify lensing events due to point-like as well as extended lenses, as well as binary lenses and variable stars such as RR Lyrae. 

Importantly, we demonstrate that this technique has potential for an early detection system that could predict transient events before they reach their peak brightness. Early prediction is critical for timely follow-up observations and for maximizing the scientific yield of detected events. We propose a framework for such a system, leveraging early identification to optimise resource allocation for follow-up studies.

This paper is organized as follows. In~\cref{sec:light curves} we discuss the sensitivity of LSST to microlensing phenomena, we introduce the classes of transients that we will be discussing throughout this paper, and we present the details of the simulated data generated for our analysis. In~\cref{sec:detection} we discuss the detection prospects of microlensing events using machine learning. This study is broken down into three analyses: the first uses an anomaly detection model to identify promising light curves and is presented in~\cref{sec:outlier-detection}; next, in~\cref{sec:early-detection}, we perform a drip-feed analysis to assess how the anomaly detection model could be used as an online alarm to help brokers identify potential microlensing events; the final analysis, presented in~\cref{sec:lsst-analysis}, employs a classifier to further help identify promising light curves that have passed the anomaly detection alarm, this will allow us to discuss the expected LSST sensitivity to different classes of transients.  In~\cref{sec:conclusions} we conclude and discuss the prospects of the potential detection of extended objects by the LSST.

\section{Light curves~\label{sec:light curves}}

\subsection{Light curve classes}
We generate light curves for five objects, including transients and RR Lyrae variables stars. In addition, we simulate a constant class (Constant) to mimic real data from non-variable sources. This class represents signal-less observations and is used to train the anomaly detection model. Due to the complexity of incorporating color information, we restrict our analysis to single-band light curves, especially the \texttt{i}-band. 

The lensing classes are comprised of light curves from point-like microlensing events (ML), binary microlensing (Binary ML), and microlensing by extended objects: NFW subhalos and boson stars (BS). These classes represent a diverse collection suitable for identifying various DM objects, including primordial black holes (PBHs), extended objects with sharply peaked density profiles (NFW subhalos), and those with flat density profiles (BS) set against a realistic astrophysical background. In this study, we neglect microlensing parallax, which is expected to be small for short duration events. In \cite{Abrams+23} it was found that parallax parameter $\pi_E$ was detected with $2 \sigma $ confidence level for a small fraction of events for $t_E \sim 100$, on the edge of our simulation range. We note that parallax can provide information about the proper motion and distance of the lens, allowing for a more precise measurement of the mass of the lens which may be explored in future work.

\subsection{Dataset generation}
For the Constant, variable, and point-like microlensing events classes, we utilise the simulation module provided in \texttt{MicroLIA}~\cite{godines2019machine}, while binary lenses are simulated using \texttt{pyLIMA}, an open-source software for microlensing analysis and simulation~\cite{Bachelet+17}. \texttt{MicroLIA} simulates variable stars using \texttt{Gatspy}'s template-based fitting method to model variable stars using real RRLyrae templates (\cite{VanderPlas+15, VanderPlasSoftware+16}).
For the extended lenses, NFW and BS, we compute the magnification using a method introduced in~\cite{Croon:2020wpr,Croon:2020ouk} with the code used in~\cite{CrispimRomao:2024nbr}.
Single-lens events are characterised by three key parameters that describe the lens-source approach: the Einstein crossing time ($t_E$), the time of maximum magnification ($t_0$), and the minimum impact parameter ($u_0$), which quantifies the separation between source and lens at $t = t_0$. The time $t_0$ is randomly drawn between the 1st and 99th percentiles of the observation window, extended by half the Einstein crossing time on either side to allow for values slightly outside the observed data range. However, this may result in signal-less simulations if $t_0$ falls within a period when the point in the sky was not visited. 

The blending fraction, $F_s$, representing the fraction of the object's flux affected by lensing, is given by 
\begin{equation}
  F_s = \frac{F_{\rm source}}{F_{\rm baseline}},
\end{equation}
and is randomly between 0 and 1, where $F_s=0$ corresponds to a scenario where the source is significantly blended.

In the case of static binary lenses where a secondary body is present, the light curve can be characterised by four additional parameters. These parameters include lens separation $s$, mass ratio $q$, source size $\rho$, and source trajectory angle $\alpha$, which determines how the source crosses the caustic structures (for a detailed discussion, see \cite{Dominik_1998a}). For the BS and NFW subhalos, we adopt the light curves described in \cite{CrispimRomao:2024nbr}, using the same ranges for the single-lens parameters listed in~\cref{tab:params}. Additionally, we include the lens size normalized to the Einstein radius, defined as $\tau_m \equiv r_{\rm lens}/r_E$, which is sampled over the range $0.05 < \tau_m < 5$.

The above parameters set the theoretical magnification of the light curve brightness by different dark objects and variable sources. In order to study the prospect of detection of dark objects by the LSST we need to generate realistic light curves. To do so, we need to simulate the survey cadence and data acquisition pipeline. To this effect, we simulated the light curves using the \texttt{baseline\_v2.0\_10yrs} Baseline Survey Strategy, published as part of ``OpSims'' and included in \texttt{rubin-sim}. We extract the 5$\sigma$ depth photometry for the `i' band using 30-second exposures, and a zeropoint of 27.85. The observation times are position-dependent as the observation strategy is inhomogeneous across the LSST footprint. As we seek to explore classification performance with sparsely sampled photometry, we select the cadences for each simulated light curve via a random selection of sky positions. This ensures uniform coverage across the entire survey, with the bulk of the observations coming from the Deep-Wide-Fast (WFD) survey mode which will comprise 90\% of the LSST observing time. 

The dataset was split into train, validation, and test subsets with relative proportions 0.5:0.25:0.25. The training set was used for exploratory data analysis and to train the machine learning models. The validation set was used for hyperparameter optimisation (when applicable) and to produce a preliminary analysis. The test set was only used to produce the final analyses presented in the next section. The dataset and the artefacts for the trained models are available at~\cite{crispim_romao_2025_15005108}.\footnote{\href{https://zenodo.org/records/15005108}{https://zenodo.org/records/15005108}}

\begin{table}[ht]
  \centering
  \begin{tabular}{cccc}
    \hline\hline
    parameter & min & max & spacing \\
    \hline
    $t_E$ (days) & 0 & 100 & linear \\
    $u_0$ & 0 & 3 & linear \\
    $t_0$ & $t_{\mathrm{1\%}}-0.5t_E$ & $t_{\mathrm{99\%}}+0.5t_E$ & linear \\
    $\rho$$\dag$ & 0 & 0.05 & log \\
    $s$$\dag$ & 0.3 & 3 & log \\
    $q$$\dag$ & 0 & 1 & linear \\
    $\alpha$ (rad)$\dag$ & 0 & 2$\pi$ & linear \\
    $\tau_m$$\ddagger$ & 0.05 & 5 & log \\
    $F_s$ & 0 & 1 & linear \\
    \hline\hline
  \end{tabular}
  \caption{Parameters used in the simulation of the lensing classes: Einstein crossing time $t_E$, minimal impact parameter (normalised to the Einstein radius) $u_0$, peak time $t_0$, source star flux $F_s$.
    \\
    $^\ddagger$ Only used to simulate the extended dark objects: normalised dark object size $\tau_m$.
    \\
  $\dag$ These parameters define the binary-lens system: the angular radius of the source star $\rho$, the projected separation between the two lens masses $s$, the mass ratio of the binary lenses $q$, the angle between the source’s trajectory and the binary lens axis $\alpha$}.
  \label{tab:params}
\end{table}

\section{Detection prospects\label{sec:detection}}
We now study the detection prospect for the classes of light curves presented in the preceding section. First, in~\cref{sec:outlier-detection}, we demonstrate how an
anomaly detection model can be used to filter out signal-less light curves, providing a purely data-driven approach to candidate selection. Then, in~\cref{sec:early-detection}, we develop a drip-feed analysis similar to that in~\cite{godines2019machine} to assess how this filtering strategy can enable the early identification of transient events in the LSST data stream, potentially triggering follow-up observations of the source. Finally, in~\cref{sec:lsst-analysis}, we conduct a classification analysis, following the approach of~\cite{CrispimRomao:2024nbr}, to investigate what can be inferred about these events based solely on LSST data, regardless of whether follow-up observations have been performed.

\subsection{Anomaly Detection\label{sec:outlier-detection}}

The first step of our analysis aims to develop a data-driven methodology that can filter out signal-less light curves to identify the most promising candidate light curves for further study. To this effect, we will train an isolation Forest (iForest)~\cite{4781136} on signal-less light curves -- more precisely, those from the Constant class described in the previous section. 
This approach enables the iForest to identify
anomalies as light curves more likely to contain a signal, either from variable sources or lensing phenomena. 

An iForest is a collection of $N$ trees $\{T_i\}$, where $i=1,\dots,\ N$, grown through random recursive partitions of a data sub-sample. The trees will grow until they reach a maximum depth or can no longer partition further (i.e. all leaf nodes have a single data point). Because the partitions are random, the number of nodes that a data point traverses from root to leaf nodes is an indication of how inline it is, with anomalies being identified as ``outliers'' 
which are more likely to arrive at a leaf node on a shorter path than an ``inlier'', i.e a signal-less light curve. Once the forest is grown, a score can be computed on a data point, $x$, as
\begin{equation}
    iFscore(x) = 0.5 - 2^{- \frac{\mathbb{E}[h(x)]}{c}} \ ,
\end{equation}
where $\mathbb{E}[h(x)]=\frac{1}{N}\sum_i h_i(x)$ is the average depth at which $x$ is found across all tress in the forest, with $h_i(x)$ denoting the depth of $x$ in tree $T_i$. The constant $c$ represents the average depth of a data point in a binary tree with the same maximal depth as the trees $\{T_i\}$. 
It is clear that $-0.5 \leq iFscore(x) \leq 0.5$, with
\begin{enumerate}
    \item $\mathbb{E}[h(x)]\to 0 \Rightarrow iFscore(x) \to - 0.5$, i.e. ``outliers'' (the anomalies) score negative values as they traverse fewer nodes than an ``inlier''. Data completely out-of-distribution will be immediately isolated and will score near $-0.5$,
    \item $\mathbb{E}[h(x)] \sim c \Rightarrow iFscore(x) \sim 0$, i.e. $0$ happens when the number of crossings is around the expected average and therefore is a reasonable cut-off between ``inliers'' and ``outliers'',\footnote{Notice that this does not mean that half the training set will be on each side of $0$. $c$ is not computed over the forest grown using the training data, but rather from an estimate for generic binary trees of the same depth. See~\cite{4781136} for more details.}
    \item $\mathbb{E}[h(x)]\to \inf \Rightarrow iFscore(x) \to 0.5$, i.e. ``inliers'' score positive values as they traverse more nodes than an ``outlier''). Data very close to the center of mass of the multivariate distribution will traverse the most nodes and will score near $0.5$. Notice that the trees are grown to a finite maximum depth, therefore $\mathbb{E}[h(x)]$ is in practice finite and therefore the ``inlier'' scores are expected to be positive but closer to $0$.
\end{enumerate}

In this work, the iForest will be trained solely on the signal-less Constant class.
The iForest then acts as a \emph{one-class classifier}, identifying non-constant light curves, i.e. those exhibiting any signal sufficiently distinct from signal-less Constant ones, as anomalies. This methodology is potentially sensitive to \emph{any} type of physical phenomena, not just to candidate signal classes introduced earlier in the previous section, which will be further studied in~\cref{sec:lsst-analysis}. 
\footnote{
In the early stages of our work, we also considered Cepheid variables with periods of $\sim3-60$ days, included in earlier studies~\cite{godines2019machine,CrispimRomao:2024nbr}. However, the iForest model easily isolated them because \texttt{MicroLIA} simulated them using a limited set of templates from real RRLyrae, which inherently have low periods ($\lesssim1$~day) and different light curve structure such as sawtooth-like asymmetries (e.g., \cite{McWilliam+11}). This highlights how unrepresentative simulations can significantly affect sensitivity analyses, potentially leading to overly optimistic assessments.
}
A similar methodology using iForest has been explored in the context of model agnostic New Physics searches in collider experiments~\cite{CrispimRomao:2020ucc}. We use \texttt{scikit-learn}~\cite{scikit-learn} implementation of the iForest, and we will leave all hyperparameters set to default.\footnote{Due to the lack of a well-defined metric for hyperparameter optimisation in semi-supervised outlier detection, we leave hyperparameters at their default values. The impact of the hyperparameters on the sensitivity to different signals is left for future work.}

\begin{figure}[ht]
  \centering
  \includegraphics[width=0.4\textwidth,trim={20cm 0 0 0},clip]{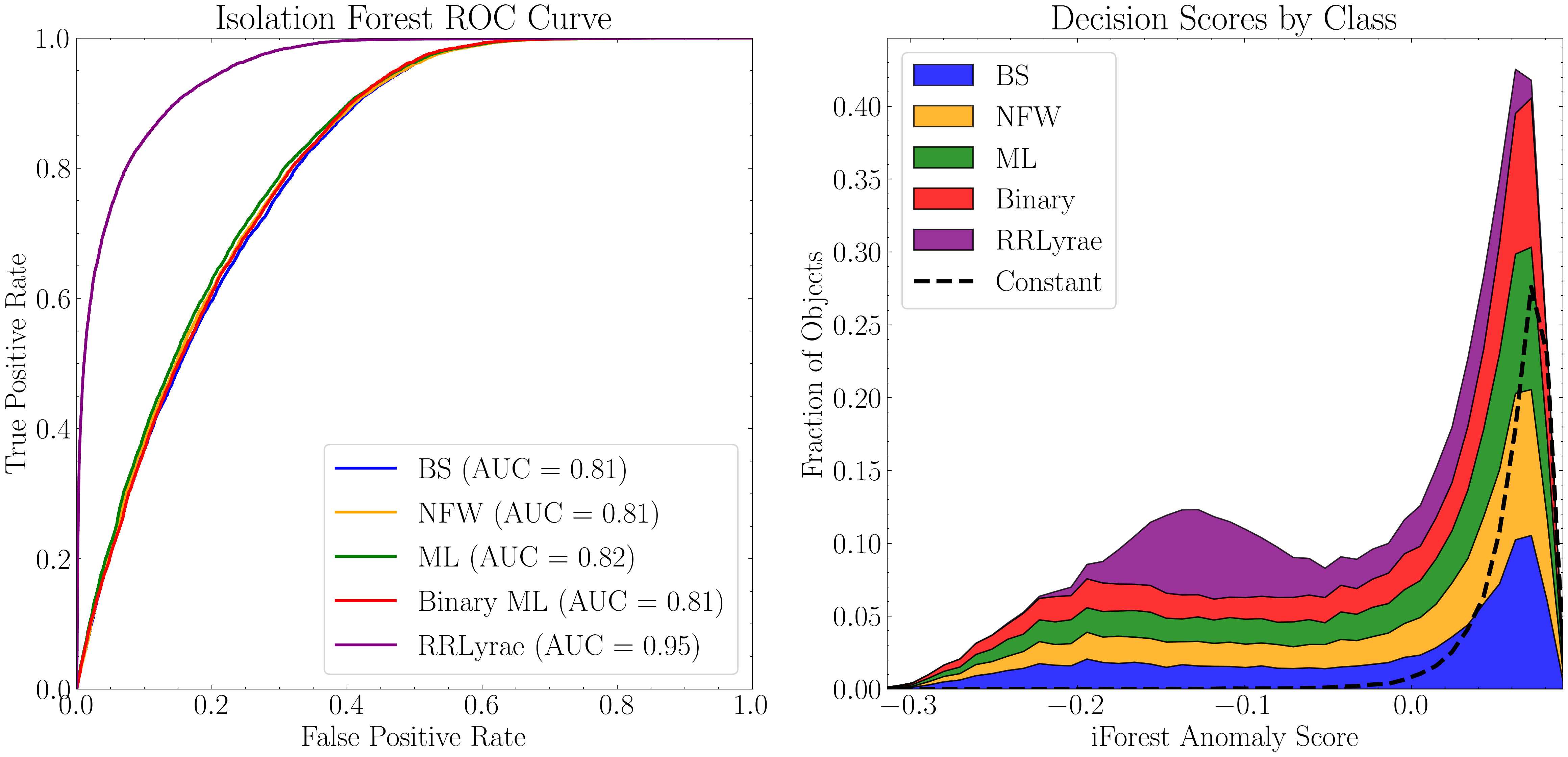}\quad
  \caption{
  Output of the iForest for the different classes.
  }
  \label{fig:iso_forest_results}
\end{figure}

In~\cref{fig:iso_forest_results}, we present the histogram of the iForest anomaly scores predictions
for different lensing classes (ML, NFW, Binary, and BS), RRLyrae variables, and Constant (signal-less) light curves.
Recalling that the simulated light curves of lensing classes might not feature a signal if the event took place outside of a visiting window, we can observe that iForest output distributions for constant and lensing classes reveal a mode of lensing light curves corresponding to cases without detectable signals during data acquisition. Additionally, we see how the values of the iForest anomaly scores for the RRLyrae are very negative, making this class especially easy to
be captured by the iForest filter. This is easy to understand: RRLyrae are variable sources which the magnitude varies for large periods of time, producing light curves distinctively different from signal-less Constant light curves, with only a small fraction not exhibiting a potential signal. From the constant test set, we find that $98\%$ of objects are assigned iForest scores $\geq 0$. Among BS objects, $45\%$ have scores $<0$, while for NFW, ML, and Binary classes, this fraction is $45\%$, $46\%$, and $46\%$, respectively. For RRLyrae variables, $71\%$ of objects have scores $< 0$.

\begin{figure}[ht]
    \centering
    \includegraphics[width=0.5\textwidth]{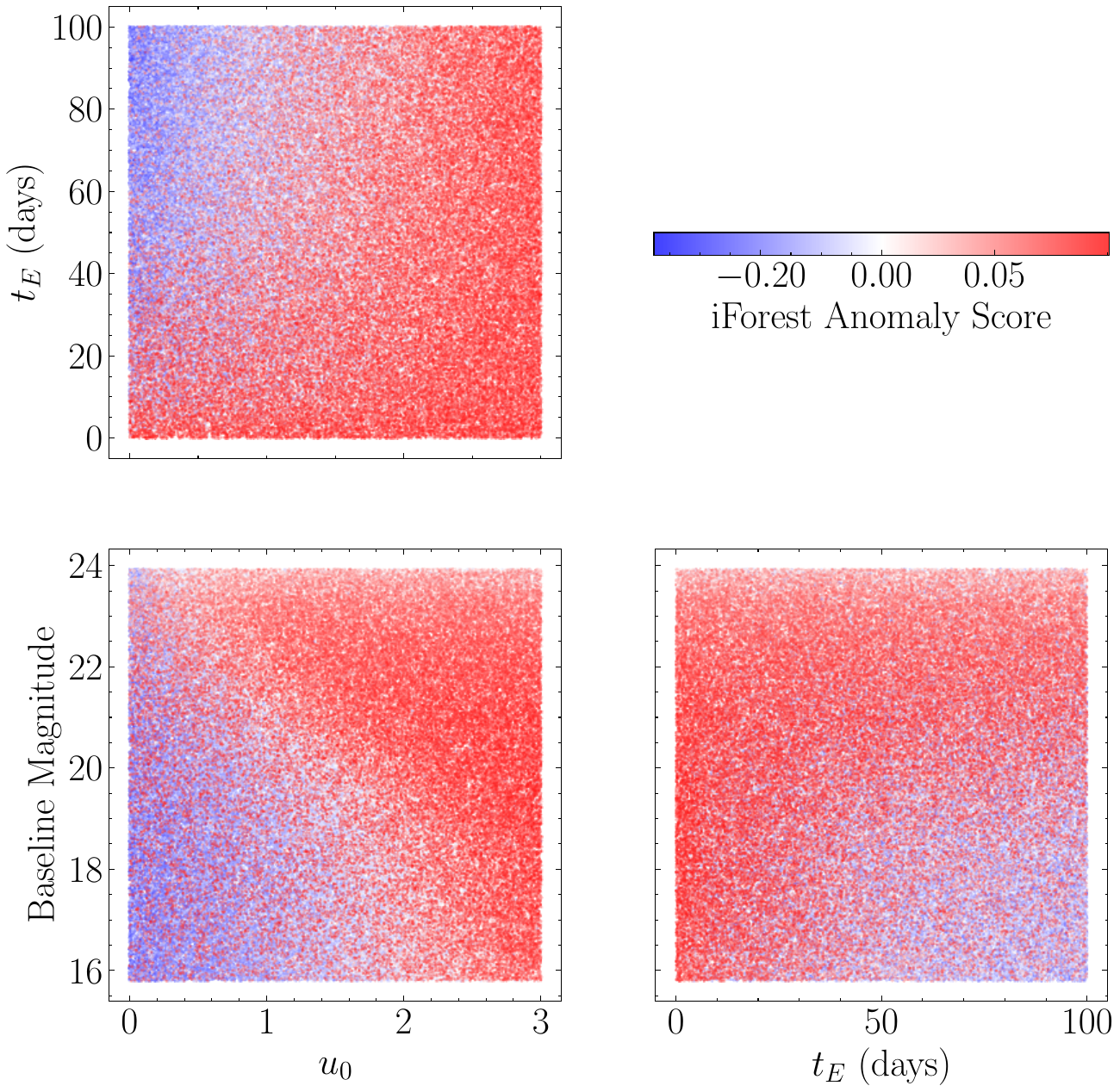}
    \caption{iForest Anomaly Score dependency on astrophysical lensing parameters: minimal impact parameter, $u_0$, Einstein crossing time, $t_E$, and source baseline magnitude. Blue (red) represents negative (positive) iForest anomaly score.}
    \label{fig:parameters_vs_iforest}
\end{figure}

These results highlight the iForest effectiveness in filtering signal-less light curves in a data-driven methodology. However, it is crucial to examine the model sensitivity to different astrophysical lensing parameters. In~\cref{fig:parameters_vs_iforest} we show the impact of the minimal impact parameter, $u_0$, the lensing event time, $t_E$, and the baseline magnitude of the source on the prediction of iForest, where blue (red) represents negative (positive) iForest scores.\footnote{We notice that there is a visual bias towards positive values of the iForest anomaly score as these points are drawn last and therefore on top.} As expected, the iForest is more sensitive to lensing events with small minimal impact parameter, $u_0 \lesssim 1.5$, which produces a higher brightness magnification; longer crossing times, $t_E \gtrsim 30$ days, which increase the likelihood of detection during observational visits; and brighter sources, magnitude $\lesssim 30$, which produces a higher signal to noise ratio. We emphasize that these parameters were not used when training the iForest, which was trained only on light curve-derived statistics.

So far, we have explored iForest as a data-driven tool for filtering signal-less light curves, enabling the identification of promising signal-containing candidates with minimal prior assumptions. However, our analysis has been purely offline, relying on full light curves simulated over a two-year span using the \texttt{rubin-sim} \texttt{baseline\_v2.0\_10yrs} survey strategy for LSST. In the next section, we assess iForest’s feasibility as an online filter for early detection, investigating its ability to identify transient events in real time during data acquisition. Given the results in~\cref{fig:iso_forest_results}, we focus on the lensing classes, as RRLyrae light curves tend to cluster around negative iForest anomaly scores, making them easier to isolate.

\subsection{Online Early Detection\label{sec:early-detection}}

As is clear from~\cref{fig:parameters_vs_iforest}, the iForest appears particularly sensitive to lensing phenomena when the minimal impact parameter is low, the Einstein crossing time is long, and the source is bright. To demonstrate its sensitivity to signal-containing light curves, we focus on this region of the lensing parameter space. For this purpose, we define a \emph{sensitivity flag} that selects light curves satisfying the following conditions:
\begin{equation}\label{eq:sensitivity_flag}
    \{u_0 < 1\ \wedge\  t_E > 40 \text{ days}\ \wedge\ \text{Baseline Magnitude} < 19\} \ ,
\end{equation}
where the values can be adjusted to increase parameter space coverage. This flag serves to illustrate iForest potential in detecting interesting light curves, which becomes evident in this region.

\begin{figure}[ht]
    \centering
    \includegraphics[width=0.5\textwidth]{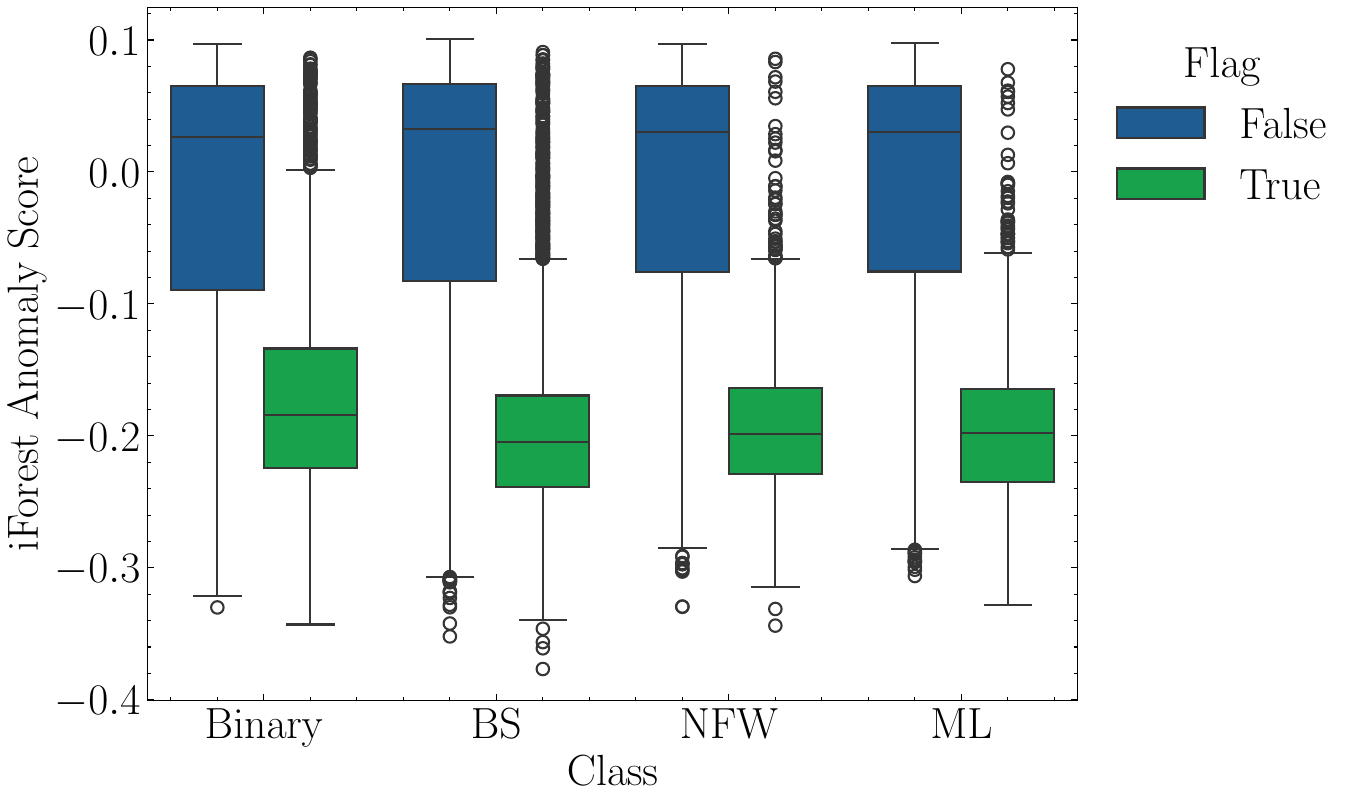}
    \caption{iForest Anomaly Score distribution for the four lensing classes, further subdivided on whether they fall in the region~\cref{eq:sensitivity_flag} or not.}
    \label{fig:lensing-classes-sensitivity-flag}
\end{figure}

In~\cref{fig:lensing-classes-sensitivity-flag}, we show the distribution of iForest anomaly scores across the four lensing classes, further subdivided by whether the light curves fall within the region defined by~\cref{eq:sensitivity_flag}. The vast majority of lensing light curves in this sensitivity region have negative iForest anomaly scores, confirming iForest’s heightened sensitivity in this subset of the parameter space. However, some light curves still yield positive scores, suggesting that even within this region certain events may appear signal-less -- likely because their observational coverage does not align with the timing of the transient event and therefore no brightness magnification is observed in the light curve. 
Additionally, we observe that BS, NFW, and ML light curves have similar distributions, whereas Binary light curves exhibit less negative iForest anomaly scores. The former observation reflects the fact that BS, NFW, and ML light curves, which are single-lens sources, resemble each other more closely than they do Binary light curves, a pattern explored further in~\cref{sec:lsst-analysis}. The latter can be attributed to the larger parameter space for Binary events compared to BS, NFW, and ML (see~\cref{tab:params}), which dilutes the number of signal-containing Binary light curves.

To evaluate the early detection capabilities of our classification engine, we conduct a drip-feeding analysis similar to the one presented in~\cite{godines2019machine}, in which the test light curves are classified one timestamp at a time. This approach effectively simulates real-time classifier performance as it ingests new data points and updates iForest predictions incrementally. Since too few data points can lead to unstable statistics, we must determine the optimal minimum number of observations (i.e., sky visits) required for reliable online anomaly detection.

\begin{figure}[ht]
    \centering
    \includegraphics[width=0.4\textwidth]{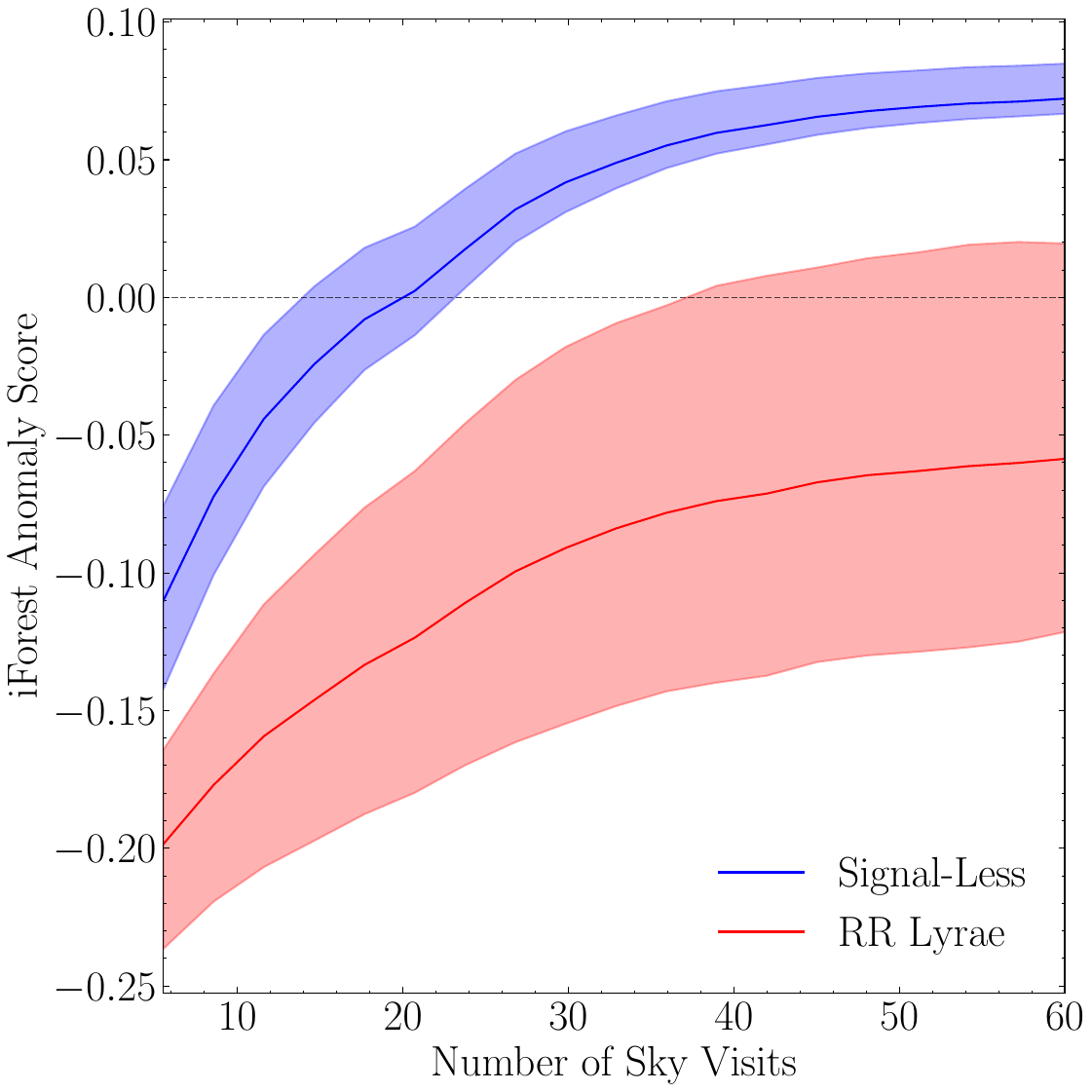}
    \caption{Comparison of anomaly scores for light curves with RR Lyrae and signal-less light curves. The solid lines show mean bin values, and the shaded regions show the 25-75 percentile range.
    }
    \label{fig:new_zp24_combined_binned_averages_RRLyrae}
\end{figure}

In~\cref{fig:main}, we illustrate how the iForest anomaly score evolves with the number of sky visits. Blue represents signal-less light curves from the Constant class. We observe that when the number of visits is below approximately $20$, iForest predictions are unreliable, often yielding negative scores that falsely indicate potential signals. However, for $30$ or more visits, signal-less Constant light curves are correctly classified, suggesting $\sim20$ timestamps are required to confidently filter out signal-less light curves, but $\sim30$ to confidently capture transient phenomena. This demonstrates the difficulty in robustly detecting transient phenomena with non-uniform cadence. The seasonal observational gaps, for example, have been shown to significantly reduce the performance of machine learning models when compared to more homogenous year-round cadence (e.g., \cite{Fagin+25}). 
For $30$ or more visits, the scores for the microlensing classes stabilise at negative values around $-0.15$, indicating that these light curves are correctly identified as non-Constant, i.e. as potentially yielding signals. This also happens to the RR Lyrae light curves, which we present in~\cref{fig:new_zp24_combined_binned_averages_RRLyrae}, where we can see that the predictions stabilise for $30$ or more sky visits, with only a small portion of RR Lyrae light curves being classified as not having signal in agreement with~\cref{fig:iso_forest_results}. We do not consider the iForest output to be reliable before stabilisation, even though the scores of signal-less and RR Lyrae light curves are separated before this. Note that no cuts similar to \eqref{eq:sensitivity_flag} have been made on the RR Lyrae parameter space, such that the percentile range band is greater in this case.

\subsection{Offline Signal Classification\label{sec:lsst-analysis}}

So far, we have examined how a semi-supervised anomaly detection method based on iForest can help \emph{remove} signal-less light curves, allowing us to isolate potential signals. In this section, we investigate the types of signals that could potentially be detected in LSST after applying iForest filtering. To this end, we perform an offline classification on light curves with an iForest score lower than $0$. We exclude Constant class light curves, as the goal of this section is to evaluate the feasibility of classifying various physical phenomena.

We follow a similar analysis to that presented in~\cite{CrispimRomao:2024nbr}, where a Histogram-based Gradient Boosting Classifier (HGBC) was used. In this study, we focus only on the transient and variable phenomena introduced in~\cref{sec:light curves}, specifically BS, NFW, point-like ML, Binary ML, and RRLyrae. We employ the \texttt{scikit-learn}~\cite{scikit-learn} implementation of the HGBC and optimised its hyperparameters using \texttt{optuna}~\cite{optuna_2019}. See~\cref{sec:hyperparameter} for more details. 

\begin{figure}[h]
  \centering
  \includegraphics[scale=0.35]{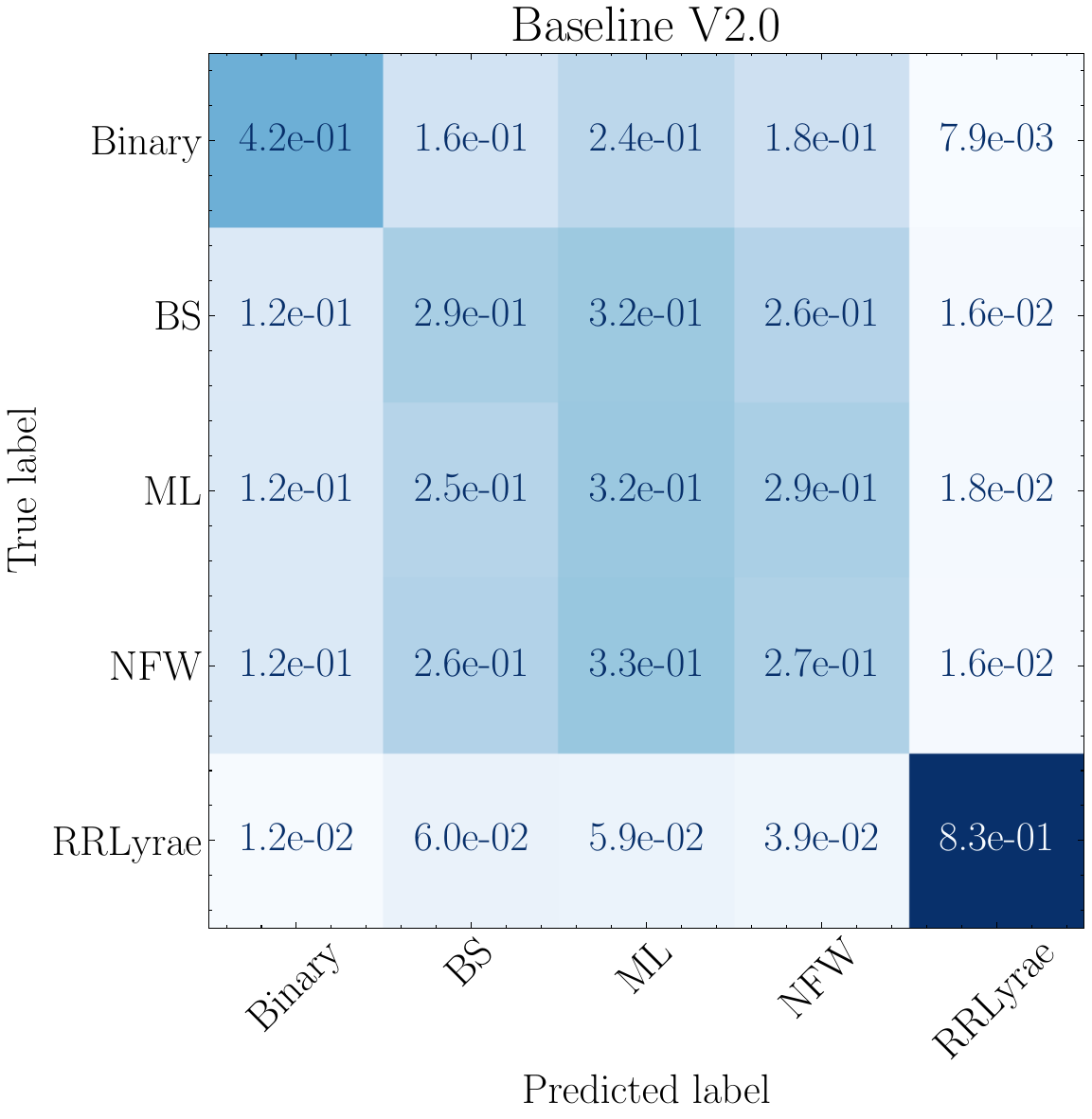}
  \caption{Confusion matrix for the classification task performed on the dataset with \texttt{rubin-sim} LSST baseline cadence simulation after performing the iForest filtering.}
  \label{fig:lsst-confusion}
\end{figure}

In~\cref{fig:lsst-confusion}, we present the confusion matrix obtained by training the HGBC on this dataset. A key observation is that microlensing events (BS, NFW, ML) are highly mixed with one another, suggesting that these classes serve as irreducible background contaminants for each other, making their separation challenging. The classifier tends to default to the point-like ML class when struggling to differentiate among these three categories. This is particularly problematic for isolating NFW light curves, as only about $27\%$ are correctly identified, with a significant fraction ($33\%$) being misclassified as ML. On a positive note, BS is apparently more distinguishable than NFW, this might indicate a potential for positive detection, which we explore further below. Additionally, Binary light curves are the easiest transient phenomena to classify, which is expected given their diverse parameter space. This results in a heterogeneous sample with distinct morphologies, making them relatively unique. However, some Binary light curves lack characteristic features and are instead classified as point-like ML. Finally, RRLyrae is the easiest class to isolate, as its light curves exhibit long-period variations absent in lensing phenomena.

\begin{figure}[h]
  \centering
  \includegraphics[scale=0.35]{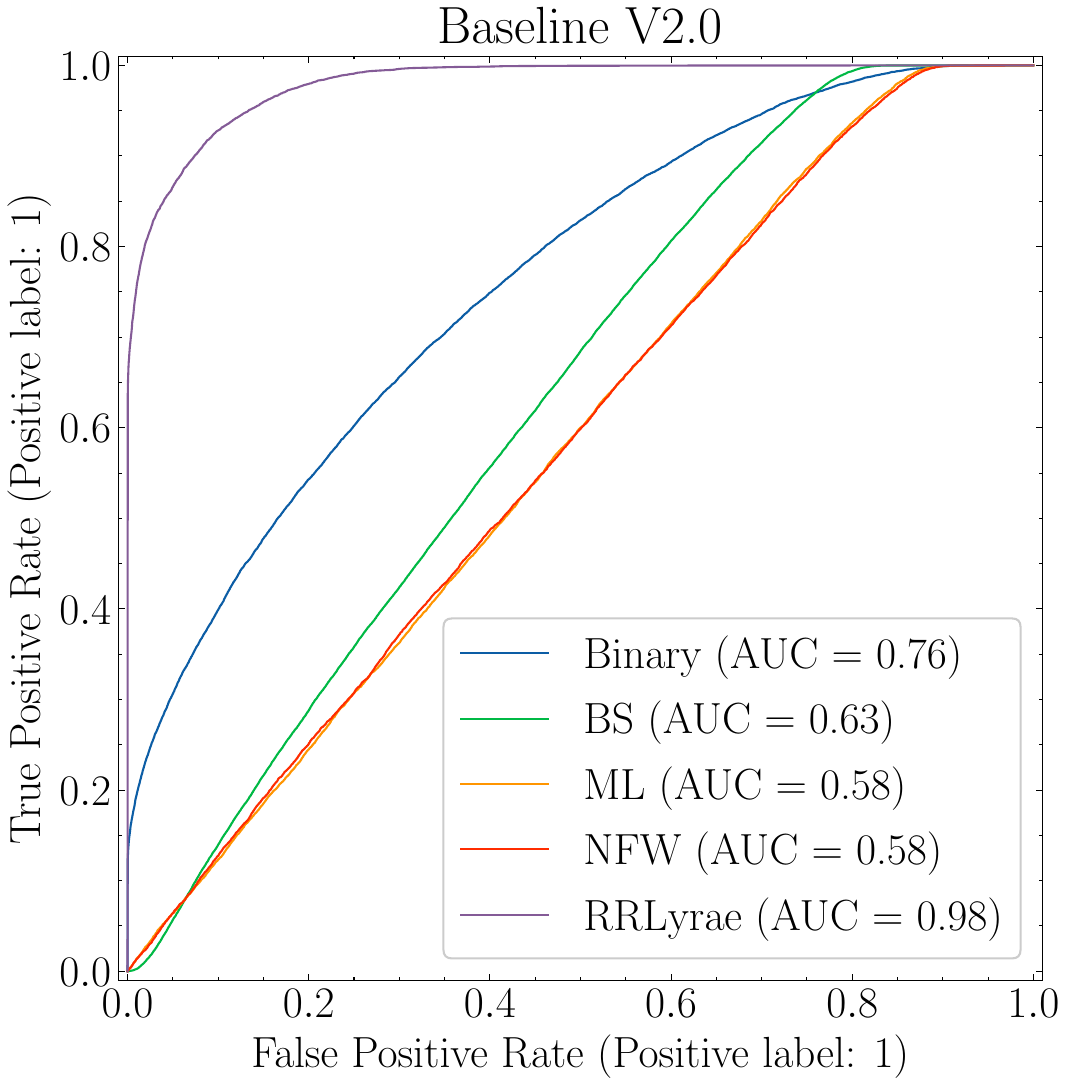}
  \caption{ROC curves and their AUC for all classes for the classification task performed on the dataset with \texttt{rubin-sim} LSST baseline cadence simulation.}
  \label{fig:lsst-roc}
\end{figure}

In~\cref{fig:lsst-roc}, we present the Receiver Operator Characteristic (ROC) curves and their corresponding Areas Under the Curve (AUC) values for the trained HGBC. We notice that the classification task is to disentangle different classes, so that the ROC AUC for each class is obtained for that class being assigned the positive label and all other classes to the negative class (one-versus-rest discrimination).

Unfortunately, we observe that for both NFW and BS, there is no regime in which the False Positive Rate vanishes while maintaining a non-zero True Positive Rate. This indicates that it is not possible to unambiguously distinguish NFW and BS light curves from point-like ML using the classifier predictions. This finding contrasts with the results of~\cite{CrispimRomao:2024nbr}, where the analysis was performed using the OGLE-II survey parameters.\footnote{Although~\cref{fig:lsst-roc} demonstrates that BS light curves are challenging to isolate, the BS ROC curve exhibits a higher AUC than those for the ML and NFW classes, suggesting that the classifier has greater discriminative power to identify BS light curves. As noted in~\cite{CrispimRomao:2024nbr}, this is due to the emergence of symmetric caustics around the main brightness peak, a distinctive feature of the BS light curves.} We attribute this difference in results to the sparse cadence of the LSST compared to the OGLE survey, which was specifically designed to detect microlensing events.

\section{Discussion\label{sec:conclusions}}
This work presents anomaly detection as a means of detecting transient events in time series data. Focussing on the LSST by the Vera C. Rubin Observatory, we have shown that training an iForest on signal-less light curves can identify transient events in all categories we simulated. This includes microlensing by DM objects: point-like DM objects, boson stars which have a flat density profile, and NFW subhalos which have an extended but peaked profile.

Importantly, our methodology can be used in a real time analysis, allowing for the early identification of transient events. This will crucially allow for follow-up strategy of the data produced by a wide-field, low cadence survey like the LSST. 

We also tested the potential of the iForest to distinguish between different classes of light curve. We found that, unlike previous studies, there is no classification regime where extended lenses can be unambiguously identified using the classifier. However, the higher sensitivity to BS hints at the possibility of a positive detection of these objects in conjugation with a more detailed analysis.

Since microlensing is fundamentally an achromatic phenomenon, our primary focus in this work is on i-band observations, which provide a clean and efficient means of detecting and characterising events without the added complexity of multi-band analysis. Color information can provide valuable insights into source properties, blending effects, and potential chromatic signatures in more complex microlensing scenarios, such as binary lenses or finite-source effects. In future work, we will incorporate multi-band photometry to explore these additional opportunities.
Moreover, while our current simulation neglects microlensing parallax -- given its expected minimal impact on the short-duration events analysed -- future iterations of our dataset will incorporate parallax effects. This will enable us to explore how parallax might help infer lens properties such as proper motion, distance, and mass, particularly in extended or binary lenses.

The anomaly detection methodology developed in this work is based on an iForest trained on an offline tabular dataset composed of statistical and derived quantities from light curves, and in~\cref{sec:outlier-detection} we showed that this ``offline'' approach is effective at finding signal-full light curves. In~\cref{sec:early-detection} we also evaluated iForest in a ``drip-feed'' scenario to assess its performance in an online detection setting. In particular, our results indicate that at least $20$ timestamps are needed before iForest predictions become reliable at rejecting signal-less light curves, highlighting a potential shortcoming in producing a timely alarm for early detection. Future work will explore alternative methods to mitigate this delay by extending our methodology with time-series anomaly detection methods and machine learning techniques designed for online learning over data streams. However, we notice that representing the data in a tabular format remains advantageous, as it condenses key features into a structured form for robust classification, while the sparse and irregular observation cadence of LSST is likely to negatively impact the performance of online methods. A detailed comparison is left for future work.

Finally, this work demonstrates the potential for anomaly detection methodology to detect transient events in the LSST data stream. 
Therefore, the primary future direction for this research program is to integrate the techniques in this work directly into LSST science brokers. This can then be applied to scheduled data releases, starting summer 2025, for offline data mining for microlensing and other transient signals. 

\section*{Acknowledgements}
We note that authorship ordering on this work is alphabetical; all authors have made important contributions to this work.
  
We thank Etienne Bachelet for providing guidance on simulating binary microlensing using \texttt{pyLIMA}, and the Rubin Observatory microlensing subgroup for useful discussions.~DC thanks the CERN theory group for hospitality during the final stages of this work.~MCR and DC are supported by the STFC under Grant No.~ST/T001011/1.

  \appendix

\section{Hyperparameter Optimisation}\label{sec:hyperparameter}

While there is no well-defined semi-supervised validation metric for tuning the hyperparameters of iForest, we can optimise the multiclass classifier used in the offline analysis to better distinguish light curves from different physical phenomena. We optimised the HGBC hyperparameters using \texttt{optuna}~\cite{optuna_2019}, training an HGBC for each proposed hyperparameter combination on the training set and selecting the optimal configuration based on the mean ROC AUC computed on the validation set. The range of hyperparameters explored during optimisation, along with their final optimal values, is listed in\cref{tab:hyperparameters}. We employed \texttt{optuna}'s default optimiser, conducting a maximum of 100 trials.
\begin{table}[h]
	\centering
	\begin{tabular}{ccc}
		\hline\hline
		Hyperparameter               & Range           & Optimal Value \\
		\hline
		\texttt{learning\_rate}      & $[0.01,\ 1]$    & $0.032$       \\
		\texttt{max\_iter}           & $[50,\ 200]$    & $200$         \\
		\texttt{max\_leaf\_nodes}    & $[20,\ 40]$     & $36$          \\
		\texttt{min\_samples\_leaf}  & $[10,\ 30]$     & $26$          \\
		\texttt{l2\_regularisation}  & $[10^{-4},\ 1]$ or $0$ & $0$       \\
		\texttt{n\_iter\_no\_change} & $[1,\ 20]$      & $18$          \\
		\hline\hline
	\end{tabular}
	\caption{Hyperparameter ranges and optimal values found during the hyperparameter optimisation step.}
	\label{tab:hyperparameters}
\end{table}

  \bibliography{refs}

  \end{document}